\documentclass{desyproc}
\usepackage{amsmath,bm,amssymb}
\usepackage{graphicx}
\usepackage{color}
\usepackage[sort&compress,numbers]{natbib}
\def\apj{Astrophys. J.}
\def\apjl{Astrophys. J. Lett.}
\def\apjs{Astrophys. J. Supp. Ser. }

\def\aap{Astron. Astrophys. }
\def\araa{Ann.\ Rev. Astron. Astroph. }

\def\mnras{Mon. Not. Roy. Astron. Soc. }

\def\prl{Phys. Rev. Lett.}
\def\prd{Phys. Rev. D.}

\def\cqg{Class. Quantum Grav.}
\def\nat{Nature}

\begin{document}
\title{New Aspects and Boundary Conditions of\\ Core-Collapse Supernova Theory}

\author{{\slshape Christian D. Ott$^{1,2,3}$, Evan P. O'Connor$^1$, 
Basudeb Dasgupta$^4$}\\[1ex]
{\footnotesize
$^1$TAPIR, California Institute of Technology, Pasadena, CA, USA\\
$^2$Center for Computation and Technology, Louisiana State University, Baton Rouge, LA, USA\\
$^3$Institute for the Physics and Mathematics of the Universe (IPMU), The University of Tokyo, Kashiwa, Japan
$^4$CCAPP, The Ohio State University, Columbus, OH, USA}
}
\contribID{xy}

\confID{???}  
\desyproc{???}
\acronym{Proceedings of HA$\nu$SE 2011} 

\maketitle
\vspace*{-0.7cm}

\begin{abstract}
Core-collapse supernovae are among Nature's grandest explosions. They
are powered by the energy released in gravitational collapse and
include a rich set of physical phenomena involving all fundamental
forces and many branches of physics and astrophysics.  We summarize
the current state of core-collapse supernova theory and discuss the
current set of candidate explosion mechanisms under scrutiny as
core-collapse supernova modeling is moving towards self-consistent
three-dimensional simulations. Recent work in nuclear theory and
neutron star mass and radius measurements are providing new
constraints for the nuclear equation of state. We discuss these new
developments and their impact on core-collapse supernova modeling.
Neutrino-neutrino forward scattering in the central regions of
core-collapse supernovae can lead to collective neutrino flavor
oscillations that result in swaps of electron and heavy-lepton
neutrino spectra. We review the rapid progress that is being made in
understanding these collective oscillations and their potential impact
on the core-collapse supernova explosion mechanism.
\end{abstract}

\vspace*{-0.4cm}
\section{Overview: Core-Collapse Supernova Theory}

The ultimate goal of core-collapse supernova theory is to understand
the mechanism driving supernova explosions in massive stars, connect
initial conditions to the final outcome of collapse, and make
falsifiable predictions of observable signals and explosion
features. These include neutrino, gravitational wave, and
electromagnetic signals, nucleosynthetic yields, compact remnant
masses, explosion morphologies, and pulsar kicks, spins, and magnetic
fields.

Baade and Zwicky, in their seminal 1934 article~\cite{baade:34b},
first hypothesized that a ``supernova represents the transition of an
ordinary star into a neutron star, consisting mainly of neutrons.''
This basic picture still holds today and the road to its refinement
has been, at best, meandering and bumpy.  When the nuclear fuel at the
core of a massive star is exhausted, the core becomes electron
degenerate and, upon reaching its effective Chandrasekhar mass,
undergoes dynamical collapse. Electron capture on free protons and
protons bound in heavy nuclei reduces the electron fraction ($Y_e$;
the number of electrons per baryon) and accelerates the collapse of
the inner core. When the latter reaches nuclear density,
$\rho_\mathrm{nuc} \approx 2.7\times
10^{14}\,\mathrm{g}\,\mathrm{cm}^{-3}$, the nuclear equation of state
(EOS) stiffens\footnote{The stiffening of the EOS near
  $\rho_\mathrm{nuc}$ is due to the repulsive effect of the strong
  force at small distances and $\Gamma = d\ln P / d\ln \rho$ jumps
  from $\sim$$4/3$ to $\gtrsim 2$. Neutron degeneracy, which is
  non-relativistic at bounce, only gives $\Gamma \approx 5/3$.},
leading to core bounce and the formation of the bounce shock at the
interface of inner and outer core. The shock initially rapidly
propagates out in radius and mass coordinate, but the work done to
break up infalling heavy nuclei and energy losses to neutrinos quickly
sap its might.  The shock stalls within tens of milliseconds of bounce
and turns into an accretion shock at a radius of
$\sim$$100-200\,\mathrm{km}$ \cite{bethe:90}.

Core collapse liberates $\sim$$3\times10^{53}\,\mathrm{erg} =
300\, $$\mathrm{\bf B}$\emph{ethe} of gravitational binding energy of
the neutron star, $\sim$$99\%$ of which is radiated in neutrinos over
tens of seconds.  The \emph{supernova mechanism} must revive the
stalled shock and convert $\sim$$1\%$ of the available energy into
energy of the explosion, which must happen within less than
$\sim$$0.5-1\,\mathrm{s}$ of core bounce in order to produce a typical
core-collapse supernova explosion and leave behind a neutron star with
the canonical neutron star gravitational mass of
$\sim$$1.4\,M_\odot$~\cite{lattimer:11,oconnor:11}.

The \emph{neutrino mechanism} \cite{bethewilson:85,bethe:90} for
core-collapse supernova explosions relies on the deposition of net
neutrino energy (heating $>$ cooling) in the region immediately behind
the stalled shock, heating this region and eventually leading to
explosion (for details, see the excellent discussion in
\cite{pejcha:11}). While having great appeal and being most
straightforward, given the huge release of neutrino energy in core
collapse, the simplest, spherically-symmetric form of this mechanism
fails to revive the shock in all but the lowest-mass massive stars
(O-Ne cores) \cite{thompson:03,liebendoerfer:05,buras:06a,kitaura:06}.

Indications are strong that multi-dimensional effects, principally
turbulent convective overturn and the standing-accretion-shock
instability (SASI, e.g., \cite{blondin:03,scheck:08} and references
therein) increase the efficacy of the neutrino mechanism by boosting
neutrino heating \cite{murphy:08,fernandez:09b,murphy:11,hanke:11} or,
as suggest by \cite{pejcha:11}, by reducing neutrino cooling.  
This is generally borne out by recent fully self-consistent
axisymmetric (2D) neutrino radiation-hydrodynamics simulations with an
energy-dependent treatment of neutrinos, but their detailed results
vary significantly from group to group and a clear picture has yet to
emerge. Marek \emph{et al.}~\cite{marek:09} reported the onset of explosions
in a nonrotating $11.2$-$M_\odot$ (at zero-age main sequence [ZAMS])
and in a slowly spinning $15$-$M_\odot$ star, setting in at
$\sim$200~ms and $\sim$600~ms after bounce, respectively, and the
estimate explosion energies are on the lower side of what is expected
from observations. The exploding simulations used the softest variant
of the EOS by Lattimer \& Swesty~(LS)~\cite{lseos:91} and included a
stronger quasi-relativistic monopole term in the gravitational
potential. A similar, so far unpublished~\cite{janka:11priv2},
calculation of core collapse in a $11.2$-$M_\odot$ star with the
stiffer EOS of H.~Shen~\emph{et~al.}~\cite{shen:98b} also produced an
explosion while simulations with the very stiff EOS by Hillebrand \&
Wolff~\cite{hillewolff:85} did not.
Bruenn~\emph{et~al.}~\cite{bruenn:09}, also using the softest LS EOS variant
and quasi-relativistic gravity, found strong
explosions setting in within $\sim$$250\,\mathrm{ms}$ after bounce in
progenitors with ZAMS masses of ($12$, $15$, $20$, and $25$)
$M_\odot$. Suwa~\emph{et~al.}~\cite{suwa:10}, using Newtonian gravity and the
soft LS EOS variant, found early, but weak explosions in a
$13$-$M_\odot$ progenitor star. Ott~\emph{et~al.}~\cite{ott:08} and
Burrows~\emph{et~al.}~\cite{burrows:06,burrows:07a}, on the other hand, who
performed purely Newtonian calculations using the stiffer H.~Shen EOS,
did not find neutrino-driven explosions in progenitors of
$11.2-25\,M_\odot$.

Given that Nature has a way to robustly (without fine tuning) explode
at least a significant fraction, but probably most stars with ZAMS
masses of $\sim$$10-20\,M_\odot$ \cite{smith:11,smartt:09b}, the large
range of differing and sometimes disagreeing results of 2D simulations
is dissatisfactory, if not disturbing.

\emph{There are essentially three possible ways out}: \\ (\emph{1})
The neutrino mechanism, while getting much closer to being viable in
2D than in 1D, may still not be reaching its full efficacy.  In 3D, an
additional fluid motion degree of freedom is available and the nature
of turbulence changes\footnote{Provided that the turbulent cascade is
  resolved, turbulent power will cascade towards small scales in 3D
  while it cascades to large scales in 2D, which is unphysical.}. This
may allow accreting material to stay even longer in the region of net
heating, resulting in a greater heating efficiency and, thus,
potentially make the neutrino mechanism robust. Results to this effect
have been obtained by Nordhaus~\emph{et~al.}~\cite{nordhaus:10} who performed
1D, 2D, and 3D calculations with parameterized neutrino heating and
cooling using spherical Newtonian gravity and the H.~Shen EOS.  This
work confirmed the results of \cite{murphy:08} for the
1D$\rightarrow$2D case and found another big increase in efficacy when
going from 2D to 3D. However, a similar parameterized study, carried
out by Hanke~\emph{et~al.}~\cite{hanke:11}, found no significant difference
between 2D and 3D. The debate thus remains open and more work will be
needed before the final word on the neutrino mechanism can be
spoken. For this, fully self-consistent 3D simulations with reliable
energy-dependent neutrino transport will be necessary.  The first
steps towards such self-consistent 3D models have already been taken
\cite{takiwaki:11b,fryerwarren:02} and their results, while
not definite, are encouraging.

(\emph{2}) If dimensionality is not the key to robust neutrino-driven
core-collapse supernova explosions, then could there be physics
missing from current 1D and 2D simulations that, once included, could
render 2D, or perhaps even 1D, neutrino-driven explosions robust? A
key example for this are self-induced (by $\nu$-$\nu$ scattering)
collective neutrino oscillations and we will discuss their potential
effect on the neutrino mechanism in \S\ref{sec:osci}.

(\emph{3}) If 3D and/or new physics cannot save the neutrino
mechanisms, alternatives must be sought. Potential ones include the
\emph{magnetorotational mechanism} (e.g., \cite{burrows:07b}), the
\emph{acoustic mechanism} \cite{burrows:06,burrows:07a,ott:06prl}, and
the \emph{phase-transition-induced} mechanisms \cite{sagert:09}.  The
\emph{magnetorotational mechanism} requires very rapid rotation in
combination with non-linear magnetic field amplification after bounce
by the magnetorotational instability (e.g.,
\cite{obergaulinger:09,burrows:07b}). Pulsar birth spin estimates
\cite{ott:06spin} and stellar evolution calculations that take into
account magnetic fields (e.g.,
\cite{heger:05}) suggest that it may be active in no more than
$\sim$$1\%$ of massive stars that produce very energetic explosions and are
related to the hyper-energetic core-collapse supernova explosions
associated with a growing number of long gamma-ray
bursts~\cite{yoon:06,woosley:06,modjaz:11}.

The \emph{acoustic mechanism}, proposed by
\cite{burrows:06,burrows:07a}, relies on the excitation of
protoneutron star pulsations by turbulence and SASI-modulated
accretion downstreams. These pulsations reach large amplitudes at
$600-1000\,\mathrm{ms}$ after bounce and damp via the emission of
strong sound waves that steepen to secondary shocks as they propagate
down the radial density gradient in the region behind the stalled shock. They
dissipate and heat the postshock region, robustly leading to
explosions, which, however, tend to be weak and occur late. This
mechanism has not been confirmed by other groups, has been studied
only in 2D simulations, and, most importantly, \cite{weinberg:08} have
shown via non-linear perturbation theory that a parametric
instability between the main mode of pulsation and abound higher-order
modes, which are not resolved by the numerical models of
\cite{burrows:06,burrows:07a}, is likely to limit the mode amplitude
to dynamically negligible magnitudes.

The \emph{phase-transition-induced mechanism} (e.g.,
\cite{gentile:93,sagert:09}) requires a hadron-quark phase transition
occurring within the first few 100~ms after core bounce (hence, at
moderate protoneutron-star central densities). This phase transition
leads to an intermittent softening of the EOS, a short collapse phase
followed by a second bounce launching a secondary shock wave that runs
into the stalled shock and launches an explosion even in spherical
symmetry. However, the needed early onset of the phase transition
requires fine tuning of the quark EOS and leads to maximum cold
neutron star gravitational masses inconsistent with observations
\cite{demorest:10,lattimer:11}.

\vskip.2cm

In the remainder of this contribution to the proceedings of the
HAmburg Neutrinos from Supernova Explosions 2011 (HA$\nu$SE 2011)
conference, we discuss, in \S\ref{sec:eos}, new boundary conditions of
core-collapse supernova theory set by neutron star mass and radius
constraints, and, in \S\ref{sec:osci}, we summarize the recent rapid
progress made by studies considering the potential effect of
collective neutrino oscillations on the core-collapse supernova
mechanism.  In \S\ref{sec:sum}, we cricially summarize our discussion
and highlight the new frontiers of core-collapse supernova theory.

\begin{figure}[t]
\begin{center}
\includegraphics[width=0.7\columnwidth]{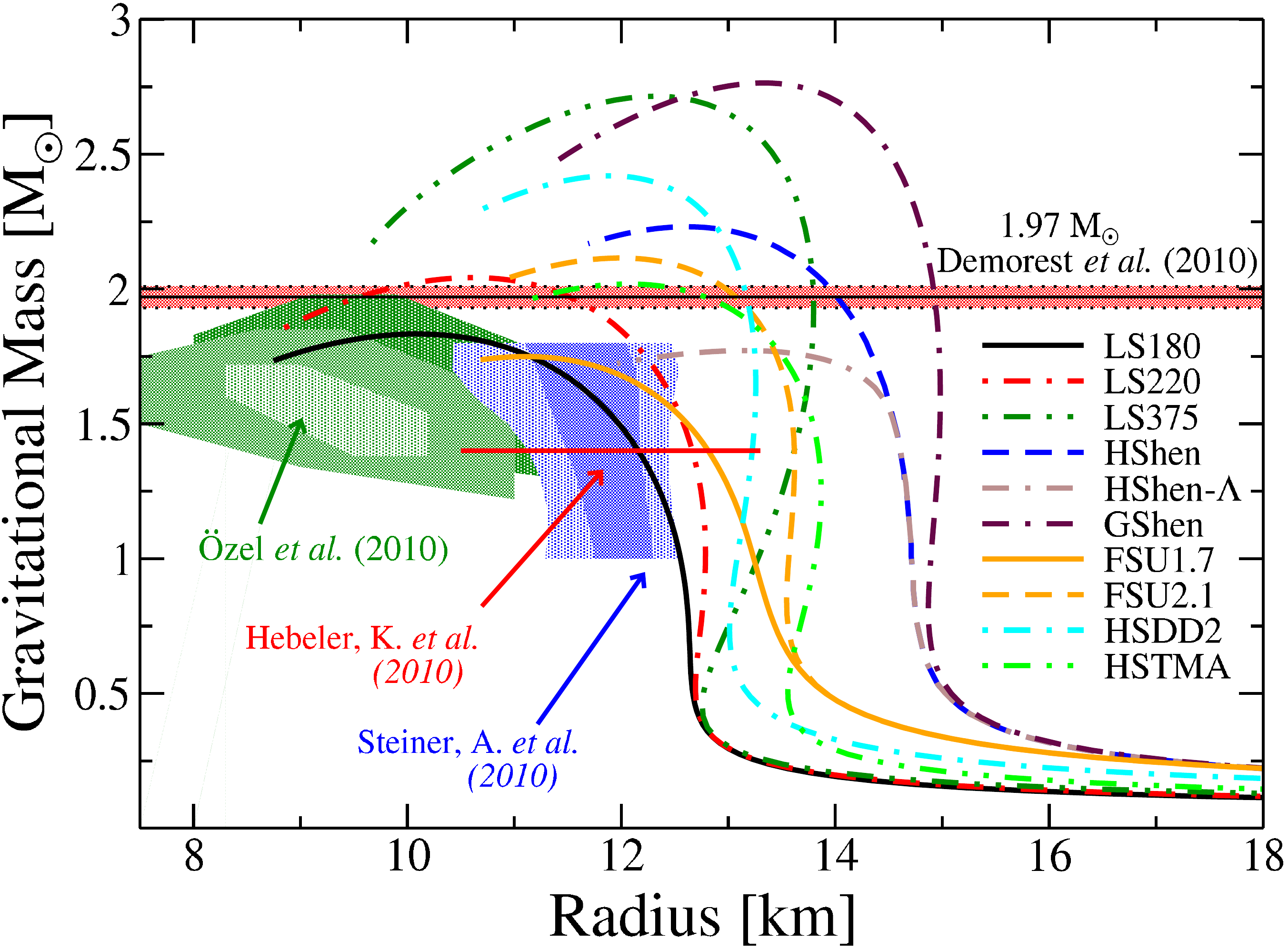} 
\caption{Mass-radius relations for 10 publically available
  finite-temperature EOS along with several constraints.  The EOS are
  taken from \cite{lseos:91, hshen:11, gshen:11a, gshen:11b,
    hempel:11, hempelweb} and the Tolman-Oppenheimer-Volkoff equation is solved
  with $T=0.1\,$MeV and neutrino-less $\beta$-equilibrium imposed.
  The family of LS EOS is based on the compressible liquid-droplet
  model~\cite{lseos:91} while all other EOS are based on relativistic
  mean field theory.  The nuclear theory constraints of Hebeler
  \emph{et al.}~\cite{hebeler:10} assume a maximum
  mass greater than $2\,M_\odot$ and do not take into account a crust
  (which would increase the radius by $\sim$$400\,\mathrm{m})$.  EOS
  that do not support a mass of at least $1.97\pm0.04\,M_\odot$ are
  ruled out \cite{demorest:10,lattimer:11}.  \"Ozel \emph{et al.}
  \cite{oezel:10b} analyzed three accreting and bursting neutron star systems 
  and derived mass-radius regions shown in
  green. Steiner \emph{et al.}~\cite{steiner:10} performed a combined
  anaylsis of six accreting neutron star systems, shown are 1-$\sigma$
  and 2-$\sigma$ results in blue.}\label{fig:MvsR_with_ranges}
\label{fig:mvsr}
\end{center}
\end{figure}

\section{New Constraints on the Supernova Equation of State}
\label{sec:eos}

An important ingredient in any core-collapse supernova model is the
nuclear EOS. It provides the crucial closure for the set of
(magneto-)hydrodynamics equations used to describe the evolution of
the collapsing stellar fluid and strongly influences the structure of
the protoneutron star and the thermodynamics of the overall problem.
Nuclear statistical equilibrium (NSE) prevails above temperatures of
$\sim$$0.5\,\mathrm{MeV}$, which corresponds to densities above
$\sim$$10^{7}\,\mathrm{g\,cm^{-3}}$ in the core-collapse supernova
problem.  In this regime, the EOS is derived from the Helmholtz free
energy and thus is expressed as a function of density $\rho$,
temperature $T$, and electron fraction $Y_e$.  The NSE part of the
core-collapse supernova EOS must cover tremendous ranges of density
($10^{7} - 10^{15}\,\mathrm{g\,cm}^{-3}$), temperature ($0.5 -
100\,\mathrm{MeV}$), and electron fraction ($0 - $$\sim$$0.6$).
Constraints from experimental nuclear physics on the nuclear EOS are
few and generally limited to only small regions of the needed
$(\rho,T,Y_e)$ space (see the discussions in \cite{hempel:11,fattoyev:10}).

A stringent constraint on the nuclear EOS is set by precision mass
measurements of neutron stars in binary systems. The $2$-$M_\odot$
($[1.97\pm 0.04]\,M_\odot$) neutron star of Demorest~\emph{et
  al.}~\cite{demorest:10} rules out a large range of soft hadronic,
mixed hadronic-exotic, and strange-quark matter
EOS~\cite{lattimer:11,oezel:10a}.

Recently, Hebeler~\emph{et al.}~\cite{hebeler:10} have carried out
chiral effective field theory calculations of neutron-rich matter
below nuclear saturation density, strongly constraining the $P(\rho)$
relationship in this regime. They derived a radius constraint for a
$1.4$-$M_\odot$ neutron star of $10.5\,\mathrm{km}\lesssim R \lesssim
13.3\,\mathrm{km}$ (these numbers would be shifted up by
$\sim$$400\,\mathrm{m}$ if a detailed crust treatment was included) by
requiring that all EOS support neutron stars with mass $\gtrsim
2\,M_\odot$ and pass through the $P(\rho)$ range allowed by their
calculations.

Steiner~\emph{et al.}~\cite{steiner:10} and \"Ozel \emph{et
  al.}~\cite{oezel:10b} analyzed observations from accreting and
bursting neutron stars to obtain neutron star mass-radius constraints.
Such observations and their interpretations should be taken with a
grain of salt, since large systematic uncertainties are attached to
the models that are required to infer mass and radius and to the
assumptions made in their statistical analysis. For example,
\cite{steiner:10} and \cite{oezel:10b}, starting with different
assumptions, derive rather different $2$-$\sigma$ mass-radius
constraints from the same set of sources.

In Fig.~\ref{fig:mvsr}, we contrast the various observational
constraints on the neutron star mass and radius with a range of
EOS used in core-collapse supernova
modeling. The LS family of EOS is based on the compressible liquid
droplet model \cite{lseos:91}, while all other EOS (drawn from
\cite{hshen:11, gshen:11a, gshen:11b, hempel:11, hempelweb}) are based
on relativistic mean field (RMF) theory. The details of the $M-R$
curves depend on multiple EOS parameters such as nuclear
incompressibility, symmetry energy and their derivatives and we must
refer the reader to \cite{hempel:11} and to the primary EOS references
for details for each EOS. Fig.~\ref{fig:mvsr} shows that none of
the current set of available EOS allow for a $2$-$M_\odot$ neutron
star while at the same time being consistent with the current mass-radius
constraints from observations.  The crux
is that the EOS needs to be sufficiently stiff to support
$2$-$M_\odot$ neutron stars \emph{and} at the same time sufficiently
soft to make neutron stars with moderate radii in the canonical mass
range. This balance appears to be difficult to realize. The stiff set
of RMF EOS produce systematically too large neutron stars. The soft
compressible liquid-droplet LS180 EOS \cite{lseos:91} agrees well with
the mass-radius constraints, but is ruled out by its failure to
support a $2$-$M_\odot$ neutron star. Closest to satisfying all
constraints are the LS220 EOS of \cite{lseos:91} and the yet unpublished HSDD2
EOS of \cite{hempelweb} based on the RMF model of \cite{typel:09}.

The stiffness of the nuclear EOS at high and intermediate densities
has important consequences for the postbounce evolution of
core-collapse supernovae.  In simple terms: the stiffer the EOS, the
more extended the protoneutron star and the larger the radius and the
lower the matter temperature at which neutrinos decouple from the
protoneutron star matter. Assuming a Fermi-Dirac spectrum with zero
degeneracy, the mean-squared energy of the emitted neutrinos is
approximately given by $\langle \epsilon_\nu^2 \rangle \approx 21
T_\nu^2$ \cite{janka:01}, where $T_\nu$ is the matter temperature (in
units of MeV) at the neutrinosphere (where the optical depth $\tau
\approx 2/3$).  Hence, a softer EOS will lead to systematically harder
neutrino spectra than a stiffer EOS (as born out by the simulations of
\cite{marek:09}). Since the charged-current neutrino heating rate
$Q_\nu^+$ scales $\propto \langle \epsilon_\nu^2 \rangle$, a soft EOS
leads to a higher neutrino heating efficiency than a stiff EOS.  This
is at least part of the explanation why some published 2D simulations
using the soft, now ruled-out LS180 EOS have shown neutrino-driven
explosions \cite{marek:09,suwa:10,bruenn:09} while simulations with
stiffer EOS have generally failed to yield such explosions in stars
more massive than $\sim$$11$ $M_\odot$
\cite{janka:11priv2,marek:09,burrows:06}.

\vspace*{-0.25cm}
\section{New Physics: Collective Neutrino Oscillations}
\label{sec:osci}

Neutrinos and antineutrinos of all three flavors are produced in
core-collapse supernovae and can oscillate from one flavor to
another. $\nu_e$ and $\bar{\nu}_e$ are made and interact via
charged-current and neutral-current interactions, while $\nu_\mu$ and
$\nu_\tau$ and their antineutrinos experience only neutral-current
processes, since no muons or tauons are present in the core-collapse
supernova environment. Hence, their interaction cross sections are
very similar and one generally lumps them together as $\nu_x =
\{\nu_\mu,\nu_\tau\}$ and $\bar{\nu}_x = \{\bar{\nu}_\mu,\bar{\nu}_\tau\}$.

\vskip.2cm

The oscillations between $\nu_e$ and $\nu_x$ or $\bar{\nu}_e$ and
$\bar{\nu}_x$ are driven by their mass differences, forward scattering off
background electrons, and forward scattering off other neutrinos and
antineutrinos. These limiting regimes are called neutrino oscillations
in vacuum~\cite{Pontecorvo:1967fh}, matter-enhanced oscillations
through the MSW effect~\cite{Mikheev:1986gs}, and collective
oscillations~\cite{Pantaleone:1992eq}, respectively. Quantitatively,
the nature of neutrino flavor conversions depends on an interplay of
vacuum neutrino oscillation frequency $\omega=\Delta m^2/(2E)$ with
the matter potential $\lambda=\sqrt{2} G_F n_e$ due to background
electrons (where $n_e$ is the electron number density) and with the
collective neutrino potential $\mu\sim\sqrt{2} G_F(1-\cos{\theta})
n_{\nu +{\bar{\nu}}}$ generated by the neutrinos themselves (where
$n_{\nu + \bar{\nu}}$ is the neutrino and antineutrino number
density).  In a typical core-collapse supernova environment, the
matter potential falls off with radius as $n_{e} \propto 1/r^3$,
whereas the collective potential falls off faster with
$n_{\nu+\bar{\nu}}\langle 1-\cos\theta\rangle\propto1/r^4$. So, when
the neutrinos travel outward from the core, they generally first
experience collective effects and then matter effects, which may be
modified by shock wave effects~\cite{Schirato:2002tg}. After they
leave the star, the mass eigenstates travel independently and are
detected on Earth as an incoherent superposition. There can be
distinctive effects due to additional conversions during propagation
inside the Earth~(e.g., \cite{Cribier:1986ak}).

\vspace*{-0.25cm}
\subsection{Collective Oscillations due to $\nu$-$\nu$ Interactions}
The neutrino density creates a potential that is not flavor diagonal
\cite{Pantaleone:1992eq}; $n_{\nu},n_{\bar{\nu}}$ are density matrices
in flavor space and depend on the flavor composition of the entire
neutrino ensemble! Flavor evolution of such dense neutrino
gases~\cite{Sigl:1992fn} can be understood to good accuracy without
considering many-particle
effects~\cite{Friedland:2003dv}. Calculations in spherical symmetry
showed that the collective oscillations can affect neutrino flavor
conversions substantially~\cite{duan:06b, duan:06a}. The main
features observed were large flavor conversions for inverted hierarchy
(neutrinos masses $m_1,m_2>m_3$), and a surprisingly weak dependence
on the mixing angle and the matter density.

These features can be understood analytically. A dense gas of
neutrinos displays collective flavor
conversion~\cite{Kostelecky:1994dt}, i.e., the flavor oscillations of
all neutrinos and antineutrinos become coupled to each other and all
of them undergo flavor conversion together. Neutrinos of all energies
oscillate almost in phase, through
synchronized~\cite{Pastor:2001iu}/parametrically
resonant~\cite{Raffelt:2008hr}/bipolar
oscillations~\cite{hannestad:06,Duan:2007mv}. The effect of the
bipolar oscillations with a decreasing collective potential $\mu$ is a
partial or complete swap of the energy spectra of two neutrino
flavors~\cite{Raffelt:2007cb, Dasgupta:2009mg}.  The
``$1-\cos\theta$'' structure of weak interactions can give rise to a
dependence of flavor evolution on the neutrino emission
angle~\cite{duan:06a} or even flavor decoherence, i.e., neutrinos
acquire uncorrelated phases, and the neutrino fluxes for all flavors
become almost identical~\cite{Raffelt:2007yz}.  For a realistic excess
of $\nu_e$, compared to $\bar{\nu}_e$ fluxes, such angle-dependent effects are
likely to be small~\cite{EstebanPretel:2007ec,Fogli:2007bk}.  Even
non-spherical source geometries can often be captured by an effective
single-angle approximation~\cite{Dasgupta:2008cu} in the coherent
regime. While most of these results were obtained for neutrino
oscillations between two flavors, it was shown that with three flavors
one can usually treat the oscillation problem {by factorizing it} into
simpler two-flavor oscillation problems, since the mass-squared
differences between the mass eigenstates obey $\Delta m_{12}^2 \ll
|\Delta m_{13}|$ and the mixing angle $\theta_{13} \ll 1$
\cite{Dasgupta:2007ws}, and the previous results are easily
generalized.
Effects of potential CP violation are expected to be small with
realistic differences between $\mu$ and $\tau$ neutrino
fluxes~\cite{Gava:2008rp}. On the other hand, similar realistic
departures are sufficient to trigger collective effects even for a
vanishing mixing angle~\cite{Blennow:2008er,Dasgupta:2010ae}.

\subsection{Results obtained with Core-Collapse Supernova Toy-Models}
\label{sec:toy}
Although the inherent nonlinearity and the presence of multi-angle
effects make the analysis rather complicated, the final outcome for
the neutrino fluxes turns out to be rather straightforward, at least
in the spherically symmetric scenario. Synchronized oscillations with
a frequency $\langle \omega \rangle$ take place just outside the
neutrinosphere at $r \sim 10-40$ km. These cause no significant flavor
conversions since the mixing angle, which determines the extent of
flavor conversion, is highly suppressed by the large matter potential
due to the high electron density in these inner regions
\cite{wolfenstein:78,mikheev:85}. A known exception occurs for the
$\nu_e$ burst phase in low-mass progenitor stars that have a very
steep density profile~\cite{Duan:2007sh}. In such a situation,
neutrinos of all energies undergo MSW resonances
\emph{before} collective effects become
negligible~\cite{Duan:2008za,Dasgupta:2008cd}. At larger radii,
$r\sim40-100$ km, bipolar or pendular oscillations
${\nu_e}\leftrightarrow {\nu_{x}}$ with a higher frequency
$\sqrt{2\omega\mu}$ follow. These oscillations are instability-driven
and thus depend logarithmically~\cite{hannestad:06} on the mixing
angle, occurring where the fluxes for the two flavors are very
similar~\cite{Dasgupta:2009mg}. As $\mu$ decreases, so that
$\langle\omega\rangle\sim\mu$, neutrinos near this instability may
relax to the lower neutrino mass (energy) state. As a result, one
finds one or more spectral swaps demarcated by sharp discontinuities
or ``spectral splits'' in the oscillated flux.

These simple explanations do not take into account the fact that
neutrinos are emitted at different angles from the neutrinosphere. As
a result, radial neutrinos take a shorter path (while tangentially
emitted neutrinos take a longer path), and thus experience less (more)
background potentials from the electrons and from other neutrinos
leading to an emission angle-dependent flavor evolution. These sort of
effects are called \emph{multi-angle effects}, and can suppress or delay 
 flavor conversions either through multi-angle matter
effects~\cite{esteban:08}, or through multi-angle neutrino-neutrino
interactions themselves~\cite{duan:11}.

\subsection{Results obtained with more realistic Models}
The observations outlined in the previous section \ref{sec:toy} were
mostly based on toy models of core-collapse supernova neutrino fluxes
and background densities. Recently, several groups have tried to
perform semi-realistic calculations of the oscillation physics, by
injecting the output neutrino fluxes from supernova simulations into
oscillation calculations~\cite{Chakraborty:2011nf, Chakraborty:2011gd,
  dasgupta:11}. Interesting results have also been obtained by
performing a linear stability analysis of the equations used for
calculating the flavor conversion, with the initial conditions taken
from simulations~\cite{Banerjee:2011fj, Sarikas:2011am}.

The simple picture given in \S\ref{sec:toy} has therefore undergone
further changes. Firstly, it has been recognized that matter effects
suppress collective oscillations even in the bipolar regime through 
multi-angle effects as explained before~\cite{Chakraborty:2011nf,
  Chakraborty:2011gd} (see also \S\ref{sec:heating}). This is most 
  effective in the pre-explosion accretion phase, when
the matter density is large in the region behind the stalled shock. In
this case, it appears that one can simply ignore collective effects
and only include the MSW effects which take place at larger radii. 
Of course,
the result depends on details of the matter density and ratios of
neutrino fluxes. In particular, for fluxes that are either highly
symmetric in neutrinos and antineutrinos~\cite{dasgupta:11}, or
include flavor dependent angular emission that leads to an angular
instability~\cite{dasgupta:11, Mirizzi:2011tu}, one finds the matter
suppression to be less effective. Secondly, in the cooling phase of
the explosion, $\nu_x$/$\bar{\nu}_x$ fluxes may be larger than
  $\nu_e$/$\bar{\nu}_e$ fluxes.
This can lead to additional instabilities
which cause multiple spectral splits~\cite{Dasgupta:2009mg}. These
features survive multi-angle effects in general, and with the
inclusion of three-flavor effects can lead to a rich and complex
phenomenology~\cite{Friedland:2010sc,Dasgupta:2010cd}.

The understanding of collective neutrino oscillations is still
evolving, and we expect that more accurate numerical calculations and
improved analytical understanding will yield new surprises and
insights into the existing results we have summarized here.

\begin{figure}[t!]
\begin{center}
\includegraphics[width=0.6\textwidth]{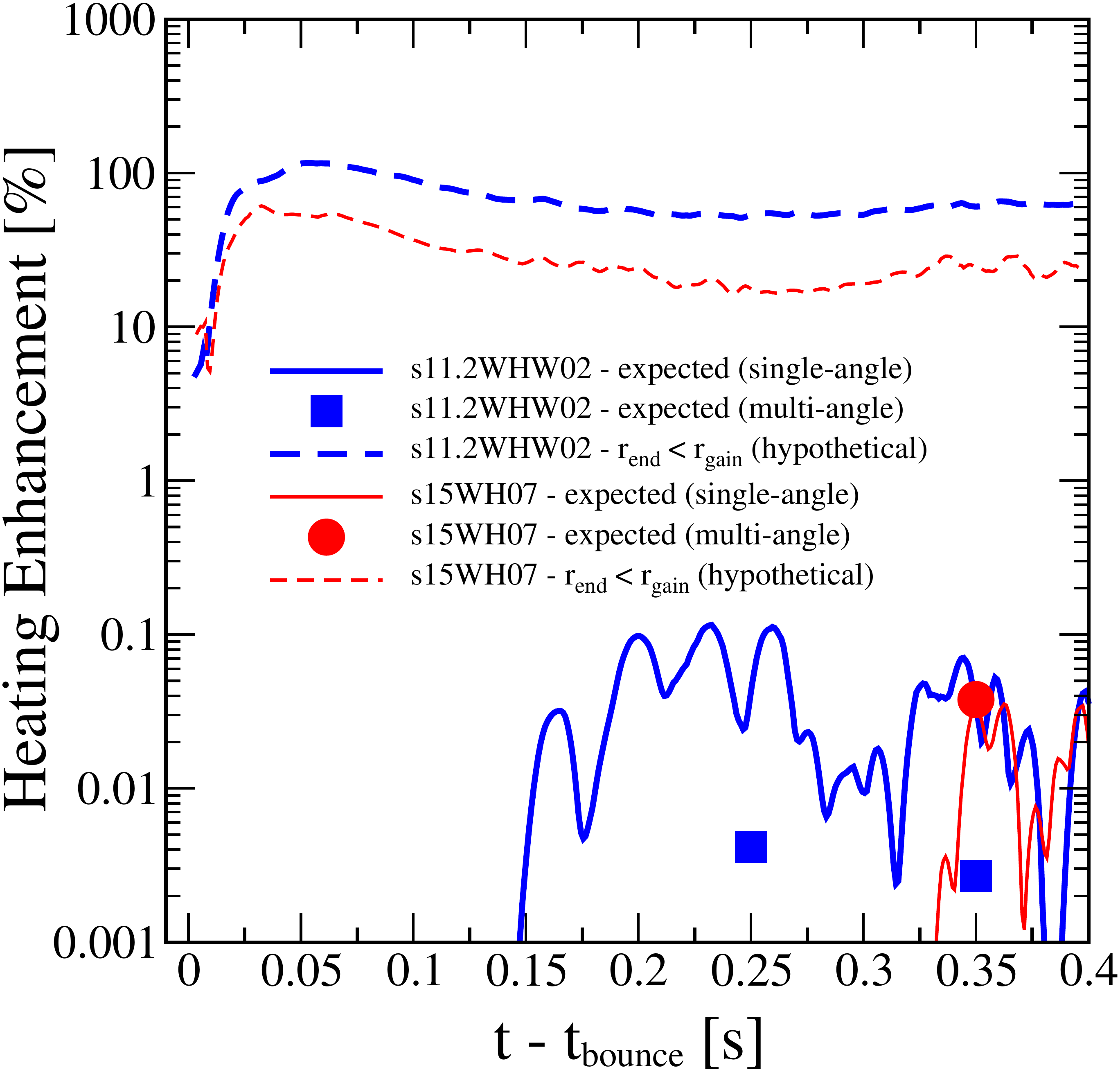} 
\caption{Time evolution (as a function of time after core bounce) of
  the potential percentage increase in the heating rate due to
  collective neutrino oscillations based on our recent simulations
  \cite{dasgupta:11}, in which we considered $11.2$-$M_\odot$ and
  $15$-$M_\odot$ progenitors.  The dashed lines assume the naive case
  of complete conversion already below the gain radius where heating
  begins to dominate over cooling. This is the case assumed by
  by Suwa~\emph{et al.}~\cite{suwa:11a}) and leads to an
  enhancement of up to $100\%$. In the our detailed oscillation
  calculations, conversion does not occur
  before the gain radius and our more realistic estimate of the
  heating enhancement is much lower
  and shown in solid lines.
  The points, blue squares for the $11.2$-$M_\odot$ model and red
  circles for the $15$-$M_\odot$ model, represent our estimate of the
  heating enhancement if multi-angle oscillation effects are included,
  which further increase the radii at which collective oscillations
  occur and thus decrease the heating enhancement even further, in
  general agreement with \cite{Chakraborty:2011nf,Chakraborty:2011gd}.
  This figure corresponds to Fig.~7 of \cite{dasgupta:11}.}
\label{fig:heating}
\end{center}
\end{figure}

\subsection{Effect of Collective Oscillations on Neutrino-driven Explosions}
\label{sec:heating}

From the core-collapse supernova theory point of view, the most
intriguing result of collective oscillations is the almost complete
exchange of $\nu_e$ and $\nu_x$ and $\bar{\nu}_e$ and $\bar{\nu}_x$
spectra in the inverted mass hierarchy.  The $\nu_x$ and $\bar{\nu}_x$
are emitted by thermal processes deep inside the core and
their spectra are much harder than those of their electron-flavor
counterparts. Due to the $\epsilon_\nu^2$-dependence of the
charged-current absorption cross section, a swap of
$\nu_x$/$\bar{\nu}_x$ and $\nu_e$/$\bar{\nu}_e$ spectra could
dramatically enhance neutrino heating and may be the crucial
ingredient missing in core-collapse supernova models, provided that
the oscillations occur at sufficiently small radii to have an effect
in the region behind the shock.  To our
knowledge, this point, in the context of collective oscillations,
 was first made by one of us \cite{ott:10jigsaw}.

Suwa~\emph{et~al.}~\cite{suwa:11a} recently performed a set of 1D and
2D core-collapse supernova simulations in which they considered
\emph{ad-hoc} spectral swaps above $9\,\mathrm{MeV}$ for neutrinos and
antineutrinos occurring at a fixed radius of $100\,\mathrm{km}$, which
is close to the gain radius (where heating begins to dominate over
cooling) in their simulations. They considered a range of progenitor
models and found that the heating enhancement by collective
oscillations can indeed turn duds into explosions.
This result was corroborated in a semi-analytic study by
Pejcha~\emph{et~al.}~\cite{pejcha:11b} in which the authors also
considered different radii for the oscillations to become effective.

Chakraborty~\emph{et~al.}
\cite{Chakraborty:2011nf,Chakraborty:2011gd} carried out the first
multi-angle single-energy neutrino oscillations based on realistic
neutrino radiation fields from 1D core-collapse supernova
simulations. They discovered that the rather high matter density
between protoneutron star and stalled shock strongly suppresses
collective neutrino oscillations in the pre-explosion phase when
multi-angle effects are taken into account. Hence, the authors excluded
any impact of collective oscillations on neutrino heating.

In Dasgupta~\emph{et al.}~\cite{dasgupta:11}, we carried out
single-angle multi-energy and multi-angle single-energy oscillation
calculations based on neutrino radiation fields from 2D core-collapse
supernova simulations performed with the {\tt VULCAN/2D}
code~\cite{burrows:07a}. In 2D, convection and SASI lead to
complicated flow patterns and large-scale shock excursions not present
in 1D simulations.  Even in our single-angle calculations and in the
most optimistic case, we find that collective oscillations do not set
in at radii sufficiently deep in the heating region to have a
significant effect on neutrino heating. When including multi-angle
effects, we also observe a suppression of collective oscillations,
though not at the level argued for by
\cite{Chakraborty:2011nf,Chakraborty:2011gd}, who made different
assumptions about the angular distribution of the neutrino radiation
fields emitted from the neutrinosphere (ours are based on the
angle-dependent neutrino transport results of \cite{ott:08}).

As depicted by Fig.~\ref{fig:heating}, we find that the heating
enhancement due to collective oscillations, if present at all, stays
below $\sim$$0.1\%$ at all times in both considered progenitor models
when oscillation radii from full oscillation calculations are taken
into account. This shows, in agreement with
\cite{Chakraborty:2011nf,Chakraborty:2011gd}, that the strong positive
effect on the neutrino mechanism reported by Suwa~\emph{et
  al.}~\cite{suwa:11a} is artificial and due primarily to their ad-hoc
choice of a small oscillation radius.

\subsection{Collective Oscillations after the Onset of Explosion}

In Dasgupta~\emph{et al.}~\cite{dasgupta:11}, we studied the suppression of
collective oscillations by multi-angle effects at high matter density
using multi-angle single-energy oscillation calculations for select
simulation snapshots of the pre-explosion phase in $11.2$-$M_\odot$
and $15$-$M_\odot$ progenitors. We also, in a more heuristic approach,
studied the potential for suppression of collective oscillations by
comparing the MSW potential $\lambda(r)$ with the expression
$\lambda_\mathrm{MA} = 2\sqrt{2}G_F \Phi_{\nu,\bar{\nu}} (R^2_{\nu_e}
/r^2)\mathcal{F}_{-}$, where $\Phi_{\nu,\bar{\nu}}$ is the neutrino
number density at the $\nu_e$ neutrino sphere radius $R_{\nu_e}$ and
$\mathcal{F}_{-} = (\Phi_{\nu_e} - \Phi_{\bar{\nu}_e}) / (\Phi_{\nu_e}
+ \Phi_{\bar{\nu}_e} + 4\Phi_{\nu_x})$ is the relative lepton
asymmetry of the neutrinos. If $\lambda \gg \lambda_\mathrm{MA}$,
collective oscillations are suppressed \cite{esteban:08}.

In \cite{dasgupta:11}, we compared $\lambda$ and $\lambda_\mathrm{MA}$
at various pre-explosion times, radii, and spatial angular directions
in our simulations using $11.2$-$M_\odot$ and $15$-$M_\odot$
progenitors.  Based on this, we concluded, in agreement with
\cite{Chakraborty:2011nf,Chakraborty:2011gd}, that suppression of
collective oscillations is likely highly relevant in the pre-explosion
phase and must be carefully studied even in relatively low-mass
progenitors with steep density profiles such as the $11.2$-$M_\odot$
progenitor model.

The situation after the onset of explosion, however, may be quite
different:  The explosion rarefies the region behind the
expanding shock and shuts off the large $\nu_e$/$\bar{\nu}_e$ accretion
luminosity, changing the neutrino flux asymmetry. Extending our
previous results to the explosion phase, we have repeated our
simulations for the $11.2$-$M_\odot$ progenitor, but included an
additional neutrino heating term in
order to drive an early explosion.  The heating term is equivalent to
the prescription used in  \cite{murphy:08} with  $L_{\nu_e} =
L_{\bar{\nu}_e} = 0.5\times10^{52}\,$ergs/s. 

\begin{figure}[t]
\begin{center}
\includegraphics[width=\columnwidth, trim=0.0cm 0.0cm 0.0cm 10.0cm, clip]{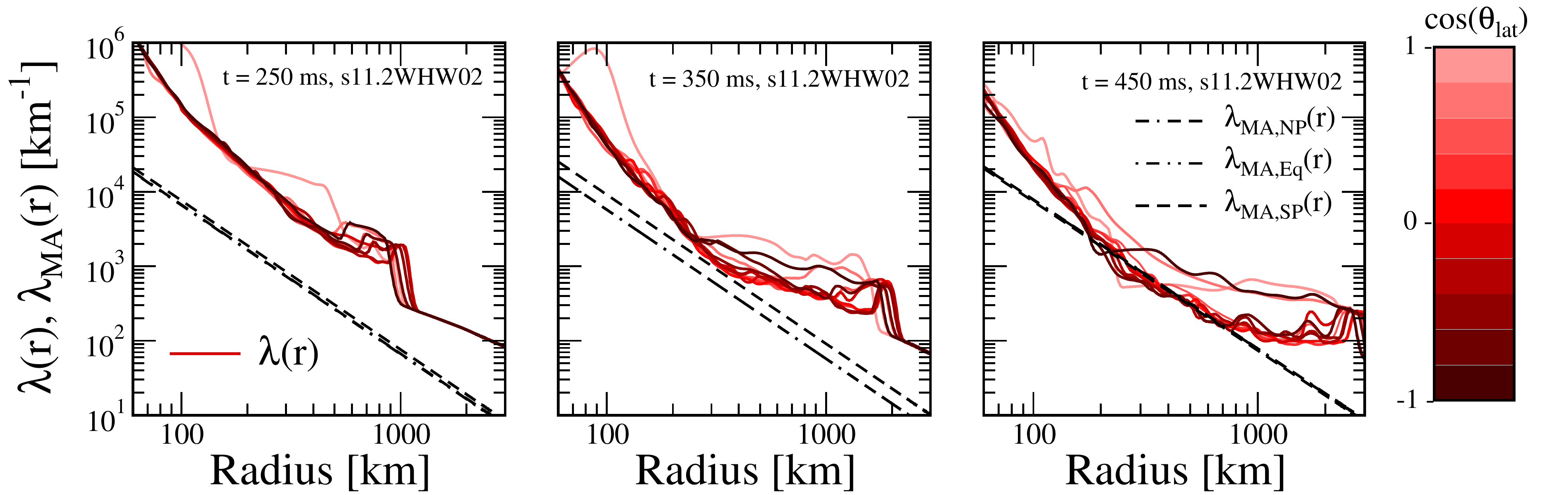} 
\caption{The MSW potential $\lambda (r)$ along various directions
  (solid lines, 10 rays equally spaced in cos[$\theta_\mathrm{lat}$]),
  in comparison to the minimum $\lambda (r)$ needed for multi-angle
  suppression, $\lambda_\mathrm{MA} =2\sqrt{2}G_F \Phi_{\nu,\bar{\nu}}
  (R^2_{\nu_e} /r^2)\mathcal{F}_{-}$. Dot-dashed-dashed,
  dot-dot-dashed, and dashed lines, indicated $\lambda_\mathrm{MA}$
  taken along the North pole (NP), equator (Eq), and South pole (SP),
  respectively. The steep rise in the $\lambda(r)$ profiles at
  $250\,$ms and $350\,$ms occurring around $r\sim 1000\,$km and $r\sim
  2000\,$km, respectively, is the location of the shock. At $450\,$ms
  the shock is close to $3000\,$km.}\label{fig:s11.2_exploding}
\end{center}
\end{figure}

In Figure \ref{fig:s11.2_exploding}, we compare the MSW potential
$\lambda(r)$ along multiple angular directions with
$\lambda_\mathrm{MA}$.  As the explosion clears out the region behind
the shock, the MSW potential decreases in strength. In this model,
within a few $100\,$ms of the onset of the explosion, the MSW
potential becomes comparable to $\lambda_\mathrm{MA}$, which indicates
that the suppression is lifted and collective oscillations may now
occur at radii as small as $\sim$$200\,\mathrm{km}$.  At this point,
the core-collapse supernova explosion has already been launched, but
collective neutrino oscillations may still affect the evolution and
various observable features, for example, via the neutrino-driven wind
from the protoneutron star and r-process
nucleosynthesis \cite{duan:11rprocess,surman:11}.

\section{Summary and Outlook}
\label{sec:sum}

In this contribution to the proceedings of the HAmburg Neutrinos from
Supernova Explosions (HA$\nu$SE) 2011 conference, we have summarized
the recent rapid progress in various aspects of core-collapse
supernova theory.  While 2D simulations continue to be perfected
\cite{ott:08,bruenn:09,mueller:10,cerda:08}, self-consistent 3D
simulations with energy-dependent neutrino radiation hydrodynamics are
now the frontier of core-collapse supernova modeling
\cite{takiwaki:11b} and are made possible by the first
generation of petascale supercomputers. General relativity is also
beginning to be included in 2D
\cite{mueller:10,cerda:08,sekiguchi:11a} and 3D simulations
\cite{ott:07prl,ott:11a}, which will eventually allow for
first-principles studies of multi-D black hole formation and the
relationship between massive star collapse and long gamma-ray bursts.
Also, open-source codes and microphysics inputs (EOS, neutrino
opacities) are gaining traction
\cite{idsaweb,stellarweb,oconnor:10,timmesweb,einsteintoolkitweb,et:11,hempelweb,hshenweb,gshenweb}. They
allow for code verification and physics benchmarking and are lowering
the technological hurdle for new groups with new ideas trying to enter
core-collapse supernova modeling.

After ten years with little activity, improved modeling capabilities,
faster computers, and the discovery of the $2$-$M_\odot$ neutron star,
have spawned a flurry of activity in the nuclear EOS community, which
has already resulted in multiple new finite-temperature EOS for 
core-collapse supernova modeling~\cite{hshen:11,gshen:11a,gshen:11b,hempel:11}.

The realization that collective neutrino oscillations may occur in the
core-collapse supernova environment
\cite{hannestad:06,duan:06a,duan:06b} has led to a plethora of work
since $\sim$2005. As we have outlined in this article, the current
state of affairs is that collective oscilations are unlikely to be
dynamically relevant in driving the explosion, but their effects are
crucial in predicting and understanding the neutrino signal that will
be seen in detectors from the next nearby core collapse event.  The
current frontier of oscillation calculations in the core-collapse
supernova context is marked by detailed multi-energy multi-angle
calculations that take their input spectra and angular distributions
from core-collapse supernova models.  Significant progress towards
this has recently been made
\cite{Chakraborty:2011gd,Chakraborty:2011nf,dasgupta:11,duan:10}, but
more will be needed to assess the potentially strong impact of
neutrino-matter interactions and only partially-decoupled neutrino
radiation fields in the oscillation regime.

The broad range of current and near-future advances in theory will be
matched and tested by observations of the next galactic (or
Magellanic-cloud) core-collapse supernova. This event will most likely
be observed in electromagnetic waves, neutrinos, and, with the
upcoming advanced generation of gravitational-wave observatories, for
the first time also in gravitational waves. Gravitational waves carry
dynamical information on the intricate multi-D processes occuring in
the supernova core \cite{ott:09,kotake:12review} and will complement
the structural and thermodynamic information carried by
neutrinos. Together, neutrinos and gravitational waves may finally
shed observational light on the details of the core-collapse supernova
mechanism.

\section{Acknowledgments}

The authors wish to thank the organizers of the HA$\nu$SE 2011
conference.  The authors furthermore acknowledge helpful discussions
with A.~Burrows, L.~Dessart, C.~Horowitz, H.-T.~Janka, J.~Lattimer, E.~Livne,
A.~Mirizzi, B.~M\"uller, J.~Murphy, J.~Nordhaus, C. Reisswig,
A.~Schwenk, G.~Shen, H.~Shen, A.~Steiner, and S.~Woosley.  CDO and EPO
are partially supported by the Sherman Fairchild Foundation and the
National Science Foundation under award numbers AST-0855535 and
OCI-0905046. Results presented in this article were obtained through
computations on the Caltech compute cluster ``Zwicky'' (NSF MRI award
No.\ PHY-0960291), on the NSF XSEDE network under grant TG-PHY100033,
on machines of the Louisiana Optical Network Initiative under grant
loni\_numrel07, and at the National Energy Research Scientific
Computing Center (NERSC), which is supported by the Office of Science
of the US Department of Energy under contract DE-AC03-76SF00098.


\begin{footnotesize}

\end{footnotesize}



\begin{thebibliography}{100}

\bibitem{baade:34b}
W.~{Baade} and F.~{Zwicky}.
\newblock {\em Proc. Nat. Acad. Sci.}, 20:259, 1934.

\bibitem{bethe:90}
H.~A. {Bethe}.
\newblock {\em Rev. Mod. Phys.}, 62:801, 1990.

\bibitem{lattimer:11}
J.~M. {Lattimer} and M.~{Prakash}.
\newblock {What a Two Solar Mass Neutron Star Really Means}.
\newblock In S.~Lee, editor, {\em From Nuclei to Stars: Festschrift in Honor of
  Gerald E. Brown. arXiv:1012.3208}. World Scientific Publishing, UK, 2011.

\bibitem{oconnor:11}
E.~{O'Connor} and C.~D. {Ott}.
\newblock {\em \apj}, 730:70, 2011.

\bibitem{bethewilson:85}
H.~A. {Bethe} and J.~R. {Wilson}.
\newblock {\em \apj}, 295:14, 1985.

\bibitem{pejcha:11}
O.~{Pejcha} and T.~A. {Thompson}.
\newblock {\em Submitted to ApJ. arXiv:1103.4865}, 2011.

\bibitem{thompson:03}
T.~A. {Thompson}, A.~{Burrows}, and P.~A. {Pinto}.
\newblock {\em \apj}, 592:434, 2003.

\bibitem{liebendoerfer:05}
M.~{Liebend{\"o}rfer}, M.~{Rampp}, H.-T. {Janka}, and A.~{Mezzacappa}.
\newblock {\em \apj}, 620:840, 2005.

\bibitem{buras:06a}
R.~{Buras}, M.~{Rampp}, H.-T. {Janka}, and K.~{Kifonidis}.
\newblock {\em Astron. Astrophys.}, 447:1049, 2006.

\bibitem{kitaura:06}
F.~S. {Kitaura}, H.-T. {Janka}, and W.~{Hillebrandt}.
\newblock {\em \aap}, 450:345, 2006.

\bibitem{blondin:03}
J.~M. {Blondin}, A.~{Mezzacappa}, and C.~{DeMarino}.
\newblock {\em \apj}, 584:971, 2003.

\bibitem{scheck:08}
L.~{Scheck}, H.-T. {Janka}, T.~{Foglizzo}, and K.~{Kifonidis}.
\newblock {\em \aap}, 477:931, 2008.

\bibitem{murphy:08}
J.~W. {Murphy} and A.~{Burrows}.
\newblock {\em \apj}, 688:1159, 2008.

\bibitem{fernandez:09b}
R.~{Fern{\'a}ndez} and C.~{Thompson}.
\newblock {\em \apj}, 703:1464, 2009.

\bibitem{murphy:11}
J.~W. {Murphy} and C.~{Meakin}.
\newblock {\em \apj}, 742:74, 2011.

\bibitem{hanke:11}
F.~{Hanke}, A.~{Marek}, B.~{M\"uller}, and H.-T. {Janka}.
\newblock {\em Submitted to the Astrophys. J., arXiv:1108.4355}, 2011.

\bibitem{marek:09}
A.~{Marek} and H.-T. {Janka}.
\newblock {\em \apj}, 694:664, 2009.

\bibitem{lseos:91}
J.~M. Lattimer and F.~D. Swesty.
\newblock {\em {Nucl. Phys. A}}, 535:331, 1991.

\bibitem{janka:11priv2}
H.-T. {Janka}.
\newblock private communication, 2011.

\bibitem{shen:98b}
H.~{Shen}, H.~{Toki}, K.~{Oyamatsu}, and K.~{Sumiyoshi}.
\newblock {\em Prog. Th. Phys.}, 100:1013, 1998.

\bibitem{hillewolff:85}
W.~{Hillebrandt} and R.~G. {Wolff}.
\newblock {Models of Type II Supernova Explosions}.
\newblock In W.~D. {Arnett} and J.~W. {Truran}, editors, {\em Nucleosynthesis :
  Challenges and New Developments}, page 131, 1985.

\bibitem{bruenn:09}
S.~W. {Bruenn}, A.~{Mezzacappa}, W.~R. {Hix}, J.~M. {Blondin}, P.~{Marronetti},
  O.~E.~B. {Messer}, C.~J. {Dirk}, and S.~{Yoshida}.
\newblock {Mechanisms of Core-Collapse Supernovae and Simulation Results from
  the CHIMERA Code}.
\newblock In G.~{Giobbi}, A.~{Tornambe}, G.~{Raimondo}, M.~{Limongi}, L.~A.
  {Antonelli}, N.~{Menci}, and E.~{Brocato}, editors, {\em AIP Phys. Conf.
  Ser.}, volume 1111 of {\em AIP Phys. Conf. Ser.}, page 593, 2009.

\bibitem{suwa:10}
Y.~{Suwa}, K.~{Kotake}, T.~{Takiwaki}, S.~C. {Whitehouse},
  M.~{Liebend{\"o}rfer}, and K.~{Sato}.
\newblock {\em Pub. Astr. Soc. Jap.}, 62:L49, 2010.

\bibitem{ott:08}
C.~D. {Ott}, A.~{Burrows}, L.~{Dessart}, and E.~{Livne}.
\newblock {\em \apj}, 685:1069, 2008.

\bibitem{burrows:06}
A.~{Burrows}, E.~{Livne}, L.~{Dessart}, C.~D. {Ott}, and J.~{Murphy}.
\newblock {\em Astrophys. J.}, 640:878, 2006.

\bibitem{burrows:07a}
A.~{Burrows}, E.~{Livne}, L.~{Dessart}, C.~D. {Ott}, and J.~{Murphy}.
\newblock {\em Astrophys. J.}, 655:416, 2007.

\bibitem{smith:11}
N.~{Smith}, W.~{Li}, A.~V. {Filippenko}, and R.~{Chornock}.
\newblock {\em \mnras}, 412:1522, 2011.

\bibitem{smartt:09b}
S.~J. {Smartt}.
\newblock {\em \araa}, 47:63, 2009.

\bibitem{nordhaus:10}
J.~{Nordhaus}, A.~{Burrows}, A.~{Almgren}, and J.~{Bell}.
\newblock {\em \apj}, 720:694, 2010.

\bibitem{takiwaki:11b}
T.~{Takiwaki}, K.~{Kotake}, and Y.~{Suwa}.
\newblock {\em Submitted to the Astrophys. J., arXiv:1108.3989}, 2011.

\bibitem{fryerwarren:02}
C.~L. {Fryer} and M.~S. {Warren}.
\newblock {\em \apjl}, 574:L65, 2002.

\bibitem{burrows:07b}
A.~{Burrows}, L.~{Dessart}, E.~{Livne}, C.~D. {Ott}, and J.~{Murphy}.
\newblock {\em \apj}, 664:416, 2007.

\bibitem{ott:06prl}
C.~D. {Ott}, A.~{Burrows}, L.~{Dessart}, and E.~{Livne}.
\newblock {\em \prl}, 96:201102, 2006.

\bibitem{sagert:09}
I.~{Sagert}, T.~{Fischer}, M.~{Hempel}, G.~{Pagliara}, J.~{Schaffner-Bielich},
  A.~{Mezzacappa}, {F.-K.} {Thielemann}, and M.~{Liebend{\"o}rfer}.
\newblock {\em Phys. Rev. Lett.}, 102:081101, 2009.

\bibitem{obergaulinger:09}
M.~{Obergaulinger}, P.~{Cerd{\'a}-Dur{\'a}n}, E.~{M{\"u}ller}, and M.~A.
  {Aloy}.
\newblock {\em \aap}, 498:241, 2009.

\bibitem{ott:06spin}
C.~D. {Ott}, A.~{Burrows}, T.~A. {Thompson}, E.~{Livne}, and R.~{Walder}.
\newblock {\em Astrophys. J. Suppl. Ser.}, 164:130, 2006.

\bibitem{heger:05}
A.~{Heger}, S.~E. {Woosley}, and H.~C. {Spruit}.
\newblock {\em \apj}, 626:350, 2005.

\bibitem{yoon:06}
S.-C. {Yoon}, N.~{Langer}, and C.~{Norman}.
\newblock {\em \aap}, 460:199, 2006.

\bibitem{woosley:06}
S.~E. {Woosley} and A.~{Heger}.
\newblock {\em Astrophys. J.}, 637:914, 2006.

\bibitem{modjaz:11}
M.~{Modjaz}.
\newblock {\em Astron. Nachr.}, 332:434, 2011.

\bibitem{weinberg:08}
N.~N. {Weinberg} and E.~{Quataert}.
\newblock {\em \mnras}, 387:L64, 2008.

\bibitem{gentile:93}
N.~A. {Gentile}, M.~B. {Aufderheide}, G.~J. {Mathews}, F.~D. {Swesty}, and
  G.~M. {Fuller}.
\newblock {\em \apj}, 414:701, 1993.

\bibitem{demorest:10}
P.~B. {Demorest}, T.~{Pennucci}, S.~M. {Ransom}, M.~S.~E. {Roberts}, and
  J.~W.~T. {Hessels}.
\newblock {\em \nat}, 467:1081, 2010.

\bibitem{hshen:11}
H.~{Shen}, H.~{Toki}, K.~{Oyamatsu}, and K.~{Sumiyoshi}.
\newblock {\em Submitted to Astrophys. J., arXiv:1105.1666}, 2011.

\bibitem{gshen:11a}
G.~{Shen}, C.~J. {Horowitz}, and S.~{Teige}.
\newblock {\em Phys.\ Rev.\ C}, 83:035802, 2011.

\bibitem{gshen:11b}
G.~{Shen}, C.~J. {Horowitz}, and E.~{O'Connor}.
\newblock {\em Phys.\ Rev.\ C}, 83:065808, 2011.

\bibitem{hempel:11}
M.~{Hempel}, T.~{Fischer}, J.~{Schaffner-Bielich}, and M.~{Liebend{\"o}rfer}.
\newblock {\em ArXiv:1108.0848}, 2011.

\bibitem{hempelweb}
M.~Hempel.
\newblock URL \url{http://phys-merger.physik.unibas.ch/~hempel/eos.html}.
\newblock Matthias Hempel's EOS webpage.

\bibitem{hebeler:10}
K.~{Hebeler}, J.~M. {Lattimer}, C.~J. {Pethick}, and A.~{Schwenk}.
\newblock {\em Phys. Rev. Lett.}, 105:161102, 2010.

\bibitem{oezel:10b}
F.~{{\"O}zel}, G.~{Baym}, and T.~{G{\"u}ver}.
\newblock {\em \prd}, 82:101301, 2010.

\bibitem{steiner:10}
A.~W. {Steiner}, J.~M. {Lattimer}, and E.~F. {Brown}.
\newblock {\em \apj}, 722:33, 2010.

\bibitem{fattoyev:10}
F.~J. {Fattoyev}, C.~J. {Horowitz}, J.~{Piekarewicz}, and G.~{Shen}.
\newblock {\em Phys. Rev. C}, 82:055803, 2010.

\bibitem{oezel:10a}
F.~{{\"O}zel}, D.~{Psaltis}, S.~{Ransom}, P.~{Demorest}, and M.~{Alford}.
\newblock {\em \apjl}, 724:L199, 2010.

\bibitem{typel:09}
S.~{Typel}, G.~{R{\"o}pke}, T.~{Kl{\"a}hn}, D.~{Blaschke}, and H.~H. {Wolter}.
\newblock {\em Phys. Rev. C}, 81:015803, 2010.

\bibitem{janka:01}
H.-T. {Janka}.
\newblock {\em \aap}, 368:527, 2001.

\bibitem{Pontecorvo:1967fh}
B.~Pontecorvo.
\newblock {\em Sov.Phys.JETP}, 26:984, 1968.

\bibitem{Mikheev:1986gs}
S.P. Mikheev and A.Yu. Smirnov.
\newblock {\em Sov.J.Nucl.Phys.}, 42:913, 1985.

\bibitem{Pantaleone:1992eq}
J.~T. Pantaleone.
\newblock {\em Phys. Lett. B.}, 287:128, 1992.

\bibitem{Schirato:2002tg}
R.~C. Schirato and G.~M. Fuller.
\newblock {\em astro-ph/0205390}, 2002.

\bibitem{Cribier:1986ak}
M.~Cribier, W.~Hampel, J.~Rich, and D.~Vignaud.
\newblock {\em Phys. Lett. B}, 182:89, 1986.

\bibitem{Sigl:1992fn}
G.~Sigl and G.~Raffelt.
\newblock {\em Nucl. Phys. B}, 406:423, 1993.

\bibitem{Friedland:2003dv}
A.~Friedland and C.~Lunardini.
\newblock {\em \prd}, 68:013007, 2003.

\bibitem{duan:06b}
H.~{Duan}, G.~M. {Fuller}, and Y.-Z. {Qian}.
\newblock {\em \prd}, 74(12):123004, 2006.

\bibitem{duan:06a}
H.~{Duan}, G.~M. {Fuller}, J.~{Carlson}, and Y.-Z. {Qian}.
\newblock {\em \prd}, 74(10):105014, 2006.

\bibitem{Kostelecky:1994dt}
V.~A. Kostelecky and S.~Samuel.
\newblock {\em \prd}, 52:621, 1995.

\bibitem{Pastor:2001iu}
S.~Pastor, G.~G. Raffelt, and D.~V. Semikoz.
\newblock {\em \prd}, 65:053011, 2002.

\bibitem{Raffelt:2008hr}
G.~G. Raffelt.
\newblock {\em \prd}, 78:125015, 2008.

\bibitem{hannestad:06}
S.~{Hannestad}, G.~G. {Raffelt}, G.~{Sigl}, and Y.~Y.~Y. {Wong}.
\newblock {\em \prd}, 74(10):105010, 2006.

\bibitem{Duan:2007mv}
H.~Duan, G.~M. Fuller, J.~Carlson, and Y.-Z. Qian.
\newblock {\em \prd}, 75:125005, 2007.

\bibitem{Raffelt:2007cb}
G.~G. Raffelt and A.~Y. Smirnov.
\newblock {\em \prd}, 76:081301, 2007.

\bibitem{Dasgupta:2009mg}
B.~Dasgupta, A.~Dighe, G.~G. Raffelt, and A.~Y. Smirnov.
\newblock {\em \prl}, 103:051105, 2009.

\bibitem{Raffelt:2007yz}
G.G. Raffelt and G.~Sigl.
\newblock {\em \prd}, 75:083002, 2007.

\bibitem{EstebanPretel:2007ec}
A.~Esteban-Pretel, S.~Pastor, R.~Tomas, G.~G. Raffelt, and G.~Sigl.
\newblock {\em \prd}, 76:125018, 2007.

\bibitem{Fogli:2007bk}
G.~L. Fogli, E.~Lisi, A.~Marrone, and A.~Mirizzi.
\newblock {\em JCAP}, 0712:010, 2007.

\bibitem{Dasgupta:2008cu}
B.~Dasgupta, A.~Dighe, A.~Mirizzi, and G.~G. Raffelt.
\newblock {\em \prd}, 78:033014, 2008.

\bibitem{Dasgupta:2007ws}
B.~Dasgupta and A.~Dighe.
\newblock {\em \prd}, 77:113002, 2008.

\bibitem{Gava:2008rp}
J.~Gava and C.~Volpe.
\newblock {\em \prd}, 78:083007, 2008.

\bibitem{Blennow:2008er}
M.~Blennow, A.~Mirizzi, and P.~D. Serpico.
\newblock {\em \prd}, 78:113004, 2008.

\bibitem{Dasgupta:2010ae}
B.~Dasgupta, G.~G. Raffelt, and I.~Tamborra.
\newblock {\em \prd}, 81:073004, 2010.

\bibitem{wolfenstein:78}
L.~{Wolfenstein}.
\newblock {\em \prd}, 17:2369, 1978.

\bibitem{mikheev:85}
S.~P. {Mikheev} and A.~Y. {Smirnov}.
\newblock {\em Yad. Fiz.}, 42:1441, 1985.

\bibitem{Duan:2007sh}
H.~Duan, G.~M. Fuller, J.~Carlson, and Y.-Z. Qian.
\newblock {\em \prl}, 100:021101, 2008.

\bibitem{Duan:2008za}
H.~Duan, G.~M. Fuller, and Y.-Z. Qian.
\newblock {\em \prd}, 77:085016, 2008.

\bibitem{Dasgupta:2008cd}
B.~Dasgupta, A.~Dighe, A.~Mirizzi, and G.~G. Raffelt.
\newblock {\em \prd}, 77:113007, 2008.

\bibitem{esteban:08}
A.~{Esteban-Pretel}, A.~{Mirizzi}, S.~{Pastor}, R.~{Tom{\`a}s}, G.~G.
  {Raffelt}, P.~D. {Serpico}, and G.~{Sigl}.
\newblock {\em \prd}, 78:085012, 2008.

\bibitem{duan:11}
H.~{Duan} and A.~{Friedland}.
\newblock {\em \prl}, 106(9):091101, 2011.

\bibitem{Chakraborty:2011nf}
S.~{Chakraborty}, T.~{Fischer}, A.~{Mirizzi}, N.~{Saviano}, and R.~{Tom{\`a}s}.
\newblock {\em Phys. Rev. Lett.}, 107:151101, 2011.

\bibitem{Chakraborty:2011gd}
S.~{Chakraborty}, T.~{Fischer}, A.~{Mirizzi}, N.~{Saviano}, and R.~{Tom{\`a}s}.
\newblock {\em \prd}, 84:025002, 2011.

\bibitem{dasgupta:11}
B.~{Dasgupta}, E.~P. {O'Connor}, and C.~D. {Ott}.
\newblock {\em Submitted to Phys. Rev. D., arXiv:1106.1167}, 2011.

\bibitem{Banerjee:2011fj}
A.~Banerjee, A.~Dighe, and G.~Raffelt.
\newblock {\em \prd}, 84:053013, 2011.

\bibitem{Sarikas:2011am}
S.~Sarikas, G.~G. Raffelt, L.~Hudepohl, and H.-T. Janka.
\newblock {\em arXiv:1109.3601}, 2011.

\bibitem{Mirizzi:2011tu}
A.~Mirizzi and P.~D. Serpico.
\newblock {\em arXiv:1110.0022}, 2011.

\bibitem{Friedland:2010sc}
A.~Friedland.
\newblock {\em \prl}, 104:191102, 2010.

\bibitem{Dasgupta:2010cd}
B.~Dasgupta, A.~Mirizzi, I.~Tamborra, and R.~Tomas.
\newblock {\em \prd}, 81:093008, 2010.

\bibitem{suwa:11a}
Y.~Suwa, K.~Kotake, T.~Takiwaki, M.~{Liebend\"orfer}, and K.~Sato.
\newblock {\em \apj}, 738:165, 2011.

\bibitem{ott:10jigsaw}
C.~D. {Ott}.
\newblock {Talk at the Joint Indo-German Supernova Astroparticle Physics
  Workshop (JIGSAW) 2010 at TIFR, Mumbai, India}, 2010.
\newblock URL \url{http://theory.tifr.res.in/~jigsaw10/talks/ott.pdf}.

\bibitem{pejcha:11b}
O.~{Pejcha}, B.~{Dasgupta}, and T.~A. {Thompson}.
\newblock {\em Submitted to the Astrophys. J., arXiv:1106.5718}, 2011.

\bibitem{duan:11rprocess}
H.~{Duan}, A.~{Friedland}, G.~C. {McLaughlin}, and R.~{Surman}.
\newblock {\em J. Phys. G Nuc. Phys.}, 38:035201, 2011.

\bibitem{surman:11}
R.~{Surman}, G.~C. {McLaughlin}, A.~{Friedland}, and H.~{Duan}.
\newblock {\em Nuc. Phys. B Proc. Suppl.}, 217:121, 2011.

\bibitem{mueller:10}
B.~{M{\"u}ller}, H.-T. {Janka}, and H.~{Dimmelmeier}.
\newblock {\em \apjs}, 189:104, 2010.

\bibitem{cerda:08}
P.~{Cerd{\'a}-Dur{\'a}n}, J.~A. {Font}, L.~{Ant{\'o}n}, and E.~{M{\"u}ller}.
\newblock {\em \aap}, 492:937, 2008.

\bibitem{sekiguchi:11a}
Y.~{Sekiguchi} and M.~{Shibata}.
\newblock {\em \apj}, 737:6, 2011.

\bibitem{ott:07prl}
C.~D. {Ott}, H.~{Dimmelmeier}, A.~{Marek}, H.-T. {Janka}, I.~{Hawke},
  B.~{Zink}, and E.~{Schnetter}.
\newblock {\em \prl}, 98:261101, 2007.

\bibitem{ott:11a}
C.~D. {Ott}, C.~{Reisswig}, E.~{Schnetter}, E.~{O'Connor}, U.~{Sperhake},
  F.~{L{\"o}ffler}, P.~{Diener}, E.~{Abdikamalov}, I.~{Hawke}, and
  A.~{Burrows}.
\newblock {\em Phys. Rev. Lett.}, 106:161103, 2011.

\bibitem{idsaweb}
M.~Liebend{\"o}rfer.
\newblock URL \url{http://www.physik.unibas.ch/~liebend/download/index.html}.
\newblock Numerical Algorithms for Supernova Dynamics.

\bibitem{stellarweb}
URL \url{http://www.stellarcollapse.org}.
\newblock {\tt stellarcollapse.org}: A Community Portal for Stellar Collapse,
  Core-Collapse Supernova and GRB Simulations.

\bibitem{oconnor:10}
E.~{O'Connor} and C.~D. {Ott}.
\newblock {\em \cqg}, 27:114103, 2010.

\bibitem{timmesweb}
F.~Timmes.
\newblock URL \url{http://cococubed.asu.edu/code_pages/codes.shtml}.
\newblock Cococubed -- Astronomy Codes.

\bibitem{einsteintoolkitweb}
URL \url{http://www.einsteintoolkit.org}.
\newblock EinsteinToolkit: A Community Toolkit for Numerical Relativity.

\bibitem{et:11}
F.~{L{\"o}ffler}, J.~{Faber}, E.~{Bentivegna}, T.~{Bode}, P.~{Diener},
  R.~{Haas}, I.~{Hinder}, B.~C. {Mundim}, C.~D. {Ott}, E.~{Schnetter},
  G.~{Allen}, M.~{Campanelli}, and P.~{Laguna}.
\newblock {\em ArXiv:1111.3344}, 2011.

\bibitem{hshenweb}
H.~Shen.
\newblock URL \url{http://physics.nankai.edu.cn/grzy/shenhong/EOS/index.html}.
\newblock Homepage of Relativistic EOS Table.

\bibitem{gshenweb}
G.~Shen.
\newblock URL \url{http://cecelia.physics.indiana.edu/gang_shen_eos/}.
\newblock Gang Shen's EOS webpage.

\bibitem{duan:10}
H.~{Duan}, G.~M. {Fuller}, and Y.-Z. {Qian}.
\newblock {\em Ann. Rev. Nuc. Part. Sc.}, 60:569, 2010.

\bibitem{ott:09}
C.~D. {Ott}.
\newblock {\em Class. Quantum Grav.}, 26:063001, 2009.

\bibitem{kotake:12review}
K.~{Kotake}.
\newblock {\em submitted to a special issue of Comptes Rendus Physique
  "Gravitational Waves (from detectors to astrophysics)", ArXiv:1110.5107},
  2011.

\end{thebibliography}
\end{document}